\documentclass[journal=jacsat,manuscript=article]{achemso}

\usepackage[version=3]{mhchem} % Formula subscripts using \ce{}
\usepackage{color}
\usepackage{amsmath}

\newcommand{\dd}{\mathrm{d}}

\newcommand{\pka}{{p$K_\mathrm{a}$}}

\author{Chenghan Li}
\author{Xing Zhang}
\author{Garnet Kin-Lic Chan}
\email{gkc1000@gmail.com}
\affiliation[Caltech]
{Division of Chemistry and Chemical Engineering, California Institute of Technology, Pasadena, USA}
\title[Quantum Alchemical]
  {General Quantum Alchemical Free Energy Simulations via Hamiltonian Interpolation}


\begin{document}

\begin{tocentry}

Some journals require a graphical entry for the Table of Contents.
This should be laid out ``print ready'' so that the sizing of the
text is correct.

Inside the \texttt{tocentry} environment, the font used is Helvetica
8\,pt, as required by \emph{Journal of the American Chemical
Society}.

The surrounding frame is 9\,cm by 3.5\,cm, which is the maximum
permitted for  \emph{Journal of the American Chemical Society}
graphical table of content entries. The box will not resize if the
content is too big: instead it will overflow the edge of the box.

This box and the associated title will always be printed on a
separate page at the end of the document.

\end{tocentry}

\begin{abstract}
We present an implementation of alchemical free energy simulations at the quantum mechanical level by directly interpolating the electronic Hamiltonian. The method is compatible with any level of electronic structure theory and requires only one quantum calculation for each molecular dynamics step in contrast to multiple energy evaluations that would be needed when interpolating the ground-state energies. We demonstrate the correctness and applicability of the technique by computing alchemical free energy changes of gas-phase molecules, with both nuclear and electron creation/annihilation. We also show an initial application to first-principles \pka~calculation for solvated molecules where we quantum mechanically annihilate a bonded proton.
\end{abstract}

\section{Introduction}
Alchemical free energy calculations\cite{chodera2011alchemical,hansen2014practical,lu2016qm, williams2018free,muegge2023recent} have emerged as a powerful tool for quantifying free energy changes associated with various physicochemical phenomena, including protein-ligand binding affinities\cite{chodera2011alchemical,rathore2013advances,williams2018free,wang2019end,lee2020alchemical,muegge2023recent}, solvation free energies\cite{abrams2006efficient,chodera2011alchemical,hansen2014practical,ostermeir2015rapid,williams2018free,procacci2019solvation,khuttan2021alchemical}, and redox potentials\cite{li2003free,zeng2008ab,zeng2009calculating,shen2018accurate}. These simulations involve an array of intermediate states that bridge the gap between two predefined endpoints of which the free energy difference is of interest. The intermediate states are ``interpolated" between the physical endpoints through a coupling parameter, typically termed $\lambda$, which continuously ``tunes" the system from one state to another.
The series of intermediates allows one to decompose the potentially large free energy into more manageable pieces that can be better sampled\cite{mey2020best}. The total free energy change is then computed using either a free energy perturbation (FEP)\cite{zwanzig1954high} or a thermodynamic integration (TI)\cite{kirkwood_theory_1968} formalism. Taking this concept further, $\lambda$-dynamics\cite{kong1996lambda,knight2009lambda,knight2011multisite}, which treats a set of $\lambda$ as extended dynamical variables evolved along with physical molecular motions, facilitates the simultaneous sampling of multiple alchemical changes (thus multiple $\lambda$'s), such as interconversion between multiple bound ligands in proteins\cite{raman2020automated,vilseck2021generalizing,robo2023fast} and between protein mutants\cite{hayes2018approaching,hayes2024sample} concurrently with the conformational space.  
$\lambda$-dynamics in the context of sampling protonation states is more commonly referred to as constant pH simulation, where the various $\lambda$ correspond to the protonation states of each protonable site\cite{lee2004constant,wallace2011continuous,donnini2011constant,swails2014constant,aho2022scalable,de2022constant,thiel2024constph}. This has been employed to simulate pH-dependent protein conformational dynamics\cite{di2012ph,chen2016conformational,yue2017constant,li2022proton}, ligand trans-membrane transport\cite{yue2019dynamic,li2022proton}, and ligand binding\cite{huang2016mechanism}.

Historically, alchemical simulations have predominantly been performed with classical molecular dynamics (MD), where molecular interactions are described by empirical energy functions, known as force fields.
While this approach has shown success in many applications, it has to be modified to work with a quantum mechanical potential energy surface (PES)\cite{yang2010qm} to simulate electronic phenomena. 
Previous work has explored two families of approaches. In the first, the quantum alchemical change is performed indirectly by employing a reference potential (usually an empirical force field) to first calculate a free energy change, which is then corrected for the discrepancies between the quantum PES and the reference potential\cite{duarte2015recent,martins2017highly,olsson2017comparison,wang2018predicting,hudson2018accelerating,giese2019development}. The primary challenge in this approach is the potentially poor overlap between the Boltzmann distribution generated by the reference potential and that from the true quantum PES. This makes accurately computing the end-point free energy correction non-trivial.
In the second, one employs an alchemical transformation that directly ``mutates" one quantum system into another.
The most common implementation computes two quantum PESs corresponding to the two end-point species and then interpolates their energies, which we will refer to as an energy interpolation or ``dual-topology" approach. This allows for general 
alchemical changes at the cost of two quantum ground-state calculations for each $\lambda$ point\cite{li2003free,li2003pka,riccardi2006development,li2007simulated}.
% the two quantum PESs are ``mechanically mixed" (herein we refer as the energy interpolation or the ``dual-topology" approach) to allow general 
% alchemical changes at the cost of two quantum ground-state calculations\cite{li2003free,li2003pka,riccardi2006development,li2007simulated}. 
Less commonly, a ``single-topology electrostatic mixing" approach has also been explored, for example, in the contexts of semi-empirical quantum methods\cite{stanton1995quantum,stanton1995general,stanton2002free}, and calculations of redox potentials\cite{zeng2008ab,shen2016quantum,von2006molecular} and solvation free energies\cite{hu2005dual}. In this case, the quantum Hamiltonian is itself interpolated (which we refer to as Hamiltonian interpolation) allowing for a single quantum ground-state calculation for each $\lambda$ point during the alchemical change.

Despite being less commonly used, the single-topology approach is more economical than the dual-topology approach in terms of the number of quantum calculations. In fact, in the case of $\lambda$ dynamics, it even has
an exponential computational advantage, because the dual-topology approach must compute $2^n$ potential energy surfaces, where $n$ is the number of $\lambda$ variables, while the single-topology approach still needs only to compute a single potential energy surface. Therefore,  in this work we revisit the single-topology approach to quantum alchemical simulations.
% in \textsc{PySCF} within both the single- and dual-topology formalisms. 
Our work goes beyond earlier investigations by supporting general alchemical changes with ab initio energy functions, 
including for methods beyond mean-field theory, and further by utilizing the $\lambda$ energy gradient, we are able to perform thermodynamic integration.
Below, we describe the formalism, our implementation, and 
proof-of-concept numerical simulations in both pure quantum and hybrid quantum mechanics/molecular mechanics (QM/MM) setups. Our implementation of both the single-topology and dual-topology approaches based on \textsc{PySCF} is available in Ref.~\cite{implementation}

\section{Theory}

We use the Born-Oppenheimer (BO) description of chemical systems where the nuclei are classical particles evolving on a PES generated by the electronic ground state. (For
% Hence, by the ``quantum free energy" we refer to the free energy of classical nuclear motion on the quantum PES generated by the electrons (but for 
a discussion of nuclear quantum effects on the free energy, see e.g. Refs.~\citenum{vanivcek2005quantum,perez2011path,habershon2011thermodynamic}). Although an important motivation is to study $\lambda$ dynamics, for simplicity and for later benchmarking with the dual-topology results, we focus here on the case of a single $\lambda$ variable, as the generalization to multiple $\lambda$ variables within the single-topology approach is straightforward. Thus we consider two chemical systems characterized by their electronic Hamiltonians, denoted $\hat{H}_0$ and $\hat{H}_1$, and the associated ground-state energies, $E_0$ and $E_1$, both of which are functions of the nuclear coordinates $\mathbf{R}$.

To facilitate the alchemical free energy calculation, we introduce a series of intermediate systems, with $\lambda$-dependent PESs, $E_\lambda$. $E_\lambda$ recovers the physical state PESs by construction i.e., it recovers $E_0$ when $\lambda = 0$, and $E_1$ when $\lambda = 1$. The free energy for a $\lambda$-dependent system is given by:
\begin{align}
    A_\lambda = -k_B T \ln{\int e^{-\beta E_\lambda} \dd\mathbf{R}}
\end{align}
where  $k_B$ is the Boltzmann constant, and $\beta$ is the inverse temperature.

Such $\lambda$-dependent states enable the breakdown of the potentially large free energy change $\Delta A=A_1-A_0$ into smaller and more manageable pieces. Depending on how the intermediate systems are used to compute $\Delta A$, there exist different flavors of alchemical simulation.
% , such as FEP, TI, and $\lambda$-dynamics. 
In this work, we focus on TI, which computes $\Delta A$ through integration of the $\lambda$ mean-gradient:
\begin{align}
    A_1 - A_0 = \int_0^1 \Big\langle \frac{\partial E_\lambda}{\partial \lambda} \Big\rangle_\lambda \dd\lambda
    \label{eq:ti}
\end{align}
where $\langle\cdot\rangle_\lambda$ indicates the average within the $E_\lambda$ ensemble.

In principle, $E_\lambda$ can be any function of $\lambda$ as long as it recovers $E_0$ and $E_1$ at the endpoints. If $E_0$ and $E_1$ are described by a force field, the interpolation of the PES can be achieved in a single-topology manner where the energy and force evaluations only involve one interpolated system with interpolated force field parameters, such as the Lennard-Jones parameters, partial charges, etc. In the quantum setting, the 
% However, such a single-topology approach is as trivial as the force field case in the quantum scenario, while a 
dual-topology approach defines an interpolated quantum PES,
% is straightforward, which 
% evolves the dynamics on the interpolated PES:
\begin{align}
    E_\lambda = (1-\lambda) E_0 + \lambda E_1 \label{eq:einterpolation}
\end{align}
i.e. the energy is interpolated. As can be seen, this requires solving for two quantum ground states at each MD step for each $E_\lambda$ ensemble.

% We refer to this scheme as the energy interpolation, and it requires solving two quantum ground states (to get individual values of $E_0$ and $E_1$) at each MD step (and requires $2^n$ quantum ground states when there are $n$ $\lambda$'s). Due to this unfavorable exponential scaling, we consider an
% In the single-topology scheme, 

It is clear from the above, that energy interpolation does not scale well with the number of $\lambda$'s (requiring $2^n$ quantum ground states when there are $n$ $\lambda$'s). This motivates the single-topology scheme of interest here that we refer to as Hamiltonian interpolation. This computes $E_\lambda$ using the ground-state energy of the interpolated electronic Hamiltonian:
\begin{align}
    \hat{H}_\lambda = (1-\lambda) \hat{H}_0 + \lambda \hat{H}_1
    \label{eq:hinterpolation}
\end{align}
Because the single-topology scheme requires only a single quantum ground state at each MD step for each $E_\lambda$ ensemble, it provides a factor of 2 savings in quantum ground-state calculations even in the case of a single $\lambda$.

We write the electronic Hamiltonian in its second-quantized form to facilitate the definition of Hamiltonian interpolation,
\begin{align}
    \hat{H} = \sum_{\mu\nu} h^\mathrm{core}_{\mu\nu} a^\dagger_\mu a_\nu + \frac{1}{2} \sum_{\mu\nu\lambda\sigma} (\mu\nu|\lambda\sigma) a^\dagger_\mu a^\dagger_\lambda a_\sigma a_\nu
\end{align}
where $\mu$, $\nu$, $\lambda$, and $\sigma$ label one-particle orbitals (for simplicity, we assume they are orthonormal). The core Hamiltonian, $h^\mathrm{core}$, is defined as:
\begin{align}
    h^\mathrm{core}_{\mu\nu} = \langle \mu | -\frac{1}{2} \nabla^2 + \hat{v}^\mathrm{ext}(\mathbf{R}) | \nu \rangle
    \label{eq:hcore}
\end{align}
and $(\mu\nu|\lambda\sigma)$ is the two-electron repulsion integral.

The Hamiltonian of two chemical systems differs only through $h^\mathrm{core}$, due to the differences in the nuclear potential $\hat{v}^\mathrm{ext}(\mathbf{R})$. The interpolation of the Hamiltonian is thus achieved by scaling the nuclear charges of atoms that are changing between the two states with the parameter $\lambda$ when computing Eq.~\ref{eq:hcore}, with
\begin{align}
    \hat{v}^\mathrm{ext}_\lambda(\mathbf{R}) = 
      \sum_{i\in 0} \frac{-(1-\lambda)Q_i}{|\mathbf{r}-\mathbf{R}_i|} 
    + \sum_{i\in 1} \frac{-\lambda Q_i}{|\mathbf{r}-\mathbf{R}_i|} 
    + \sum_{i\in\mathrm{common}} \frac{-Q_i}{|\mathbf{r}-\mathbf{R}_i|} 
\end{align}
Here, $Q_i$ is the nuclear charge of atom $i$, ``0'' and ``1'' are atom indices unique to states 0 and 1, respectively, and ``common'' represents atoms shared by both states. In a periodic QM/MM setup\cite{li2024accurate}, the difference in $\hat{H}$ still lies in only $\hat{v}^\text{ext}$, which contains the electrostatic potential from all QM and MM nuclei and their periodic images. The $\hat{v}_\lambda^\mathrm{ext}$ is given by interpolating the QM nuclear charges and, if MM atoms are subject to alchemical changes, also the interpolated MM partial charges. 

When atom-centered orbitals are used to express the Hamiltonian, the two $\hat{H}_0$ and $\hat{H}_1$ do not share the same dimension since the basis functions carried by atoms in one endpoint state may be absent in another state. We thus represent the Hamiltonian in the union basis of both endpoints and then introduce the following term to $\hat{H}_\lambda$ to penalize electron populations on ``ghost" orbitals of the absent atoms:
\begin{align}
    \hat{h}^\mathrm{orb}_\lambda = \sum_{\mu\in 1}v^\mathrm{orb} f_0(\lambda) a^\dagger_\mu a_\mu + \sum_{\mu\in 0} v^\mathrm{orb} f_1(\lambda) a^\dagger_\mu a_\mu
    \label{eq:horb}
\end{align}
where $f_0$ and $f_1$ are two smooth switching functions that are turned on/off in states 0/1 and 1/0, respectively. 
By setting $v^\mathrm{orb}$ to a sufficiently large positive value, the ``ghost" orbitals on the annihilated atoms in either state become unpopulated and thus are effectively deleted. 
In principle, $f_0$ and $f_1$ can be any smooth switching functions, and we use the following two to ensure such potential is quickly turned on only when atoms have nearly disappeared:
\begin{align}
    f_0(\lambda) & = e^{-20\lambda} \\
    f_1(\lambda) & = e^{-20(1-\lambda)} 
\end{align}
%It is worth noting that $\hat{h}^\mathrm{orb}$ is only necessary for atom-centered bases like Gaussian orbitals but not for plane waves.

The above formalism can clearly be used with any electronic structure method that can compute a ground-state potential energy surface for a given number of electrons $N^\text{elec}$.
However, in addition to a change in the Hamiltonian itself, the number of electrons can also vary in the alchemical process. To accommodate this, we choose to work in the grand canonical ensemble of electrons at a finite electronic temperature $T^\mathrm{elec}$\cite{mermin1965thermal,weinert1992fractional}(where $T^\mathrm{elec}$  is not correlated with the nuclear temperature, and if chosen non-zero is primarily used to facilitate convergence of the electronic calculations) and chemical potential $\mu$. 
When evolving the nuclear dynamics, we therefore use the electronic free energy surface instead of the ground state PES. 
% While the finite-temperature electron correlation theories are under active development, the mean-field theory has been well-established and 
The finite-temperature electronic calculation formally approximates the grand canonical potential
\begin{align}
    W^\text{elec} = E^\text{elec} - T^\text{elec} S^\text{elec} - \mu N^\text{elec}
\end{align}
which is related to the (Helmholtz) electronic free energy on which surface we propagate by 
\begin{align}
    A^\mathrm{elec}(T^\text{elec}, N^\text{elec}) = W^\text{elec} + \mu(T^\text{elec}, N^\text{elec}) N^\text{elec}.
\end{align}
where in the above, we are indicating that $\mu$ is adjusted in the grand canonical ensemble so as to give the target $N^\text{elec}$ that characterizes the free energy $A^\mathrm{elec}$ (but, strictly speaking, computed with the fluctuations of the grand canonical ensemble).
In principle, the only requirement along the $\lambda$ path is that $T^\text{elec}$ is sufficiently low, and $\mu$ is correctly adjusted so that one obtains the ground state for the correct number of electrons at each of the endpoints. 
As such, we consider not only $N^\mathrm{elec}$ to be $\lambda$-dependent but also that $T^\mathrm{elec}$ follows a $\lambda$-dependent schedule if necessary. Given a finite-temperature grand canonical electronic structure method, the $\lambda$ interpolation can then be performed in either the single- or dual-topology approaches.

Due to the loss of interactions with the rest of the system, the annihilated atoms need to be geometrically restrained to avoid sampling difficulties. Although conceptually orthogonal to the quantum alchemical transformation, we briefly discuss the restraint here for completeness. We consider a restraint potential taking the form
\begin{align}
    E^\mathrm{res}_\lambda = U_\lambda(\mathbf{R}_2-\mathbf{f}(\mathbf{R}_1))
    \label{eq:Eres}
\end{align}
where $U_\lambda$ is an arbitrary energy function that is turned off when $\lambda=0$ and is fully turned on when $\lambda=1$, such as a harmonic potential modulated by $\lambda$, $\mathbf{R}_2$ indicates the annihilated atom positions, while $\mathbf{R}_1$ is the remaining atom positions, and $\mathbf{f}$ can be any linear or nonlinear transformation on the $\mathbf{R}_1$ coordinates. (We assume that $\lambda=0$ corresponds to the state when the $\mathbf{R}_2$ atoms are present and $\lambda=1$ corresponds to the state when they are annihilated.) It is straightforward to show that the free energy change due to introducing such restraint is
\begin{align}
    \Delta A^\mathrm{res} = -k_B T \ln{(\int e^{-\beta U_1(\mathbf{R}_2)} \dd\mathbf{R}_2 / \int 1 \dd\mathbf{R}_2)}
\end{align}
which can be easily computed independently of the system PES. The free energy values reported in this work do not include this correction, which depends on the volume associated with the denominator in the above integral, and as such the values are subject to the specific restraint we adopt.

With all the above considerations, we evolve the nuclear dynamics on the following $\lambda$-dependent PES:
\begin{align}
    E_\lambda = E^\mathrm{nuc}_\lambda + A^\mathrm{elec}_\lambda + E_\lambda^\mathrm{res} + E_\lambda^\mathrm{misc}
\end{align}
where $E^\mathrm{nuc}_\lambda$ is the classical Coulomb interaction between bare nuclei, modulated by $\lambda$ through interpolated nuclear charges, 
$A^\text{elec}_\lambda$ is the electronic free energy (or zero-temperature free energy if there is no particle number change). 
% is given by Eq.~\ref{eq:einterpolation} for the dual-topology formalism and (the ground-state of)~Eq.~\ref{eq:hinterpolation} for the single-topology formalism (or their finite temperature extensions),
% and $E^\mathrm{res}_\lambda$ is given by Eq.~\ref{eq:Eres}.
$E^\mathrm{misc}_\lambda$ contains all the other energy terms, including the dispersion energy when a dispersion-corrected density functional theory is employed, and the MM electrostatic energy, which depends on $\lambda$ if MM charges are alchemically changed. The $\lambda$-dependent dispersion energy can be computed via energy interpolation (even for many $\lambda$'s) since it is computationally cheap, and the $\lambda$-dependent MM energy is computed by standard MD techniques with the interpolated MM charges. 

We now discuss the $\lambda$ gradient, as required to perform the TI (Eq.~\ref{eq:ti}) in this work. (It is also necessary to carry out the $\lambda$-dynamics which we will present in the future). The nuclear energy and restraint energy are simple force field terms and their derivatives are straightforward. The electronic free energy $A^\mathrm{elec}$ gradient has two contributions, from the $\lambda$-dependent Hamiltonian response and the electron number and temperature change. Given that the $\lambda$-dependence is confined to the one-particle Hamiltonian, the electronic (free) energy response takes a simple form:
\begin{align}
    \frac{\partial A^\mathrm{elec}_\lambda}{\partial \lambda} \text{(1st term)}
    = \sum_{\mu\nu} \Big(\frac{\partial h^\mathrm{core}_{\mu\nu}}{\partial \lambda} + \frac{\partial h^\mathrm{orb}_{\mu\nu}}{\partial \lambda}\delta_{\mu\nu}\Big) \gamma_{\mu\nu}
    \label{eq:dAdl_1}
\end{align}
In this equation, $\gamma_{\mu\nu}$ is the (relaxed) one-particle reduced density matrix (1-RDM). The $h^\mathrm{orb}$ response is straightforwardly obtained by differentiating Eq.~\ref{eq:horb}, and the gas-phase $h^\mathrm{core}$ response is computed from the one-electron integrals involving the changing atoms:
\begin{align}
    \frac{\partial h^\mathrm{core}_{\mu\nu}}{\partial \lambda} 
    = \langle \mu | \sum_{i\in0} \frac{Q_i}{|\mathbf{r}-\mathbf{R}_i|}| - \sum_{i\in1} \frac{Q_i}{|\mathbf{r}-\mathbf{R}_i|} |\nu \rangle
\end{align}
In QM/MM, $\partial h_{\mu\nu}^\mathrm{core}/\partial\lambda$ becomes more complicated due to the electrostatic interactions between the QM and MM subsystems and their periodic images, but its calculation does not require recomputing the full electrostatics, but only the interactions involving the changing charges. We refer to the implementation provided by this work for the details~\cite{implementation}, which depend on the specifics of the QM/MM approach.

The response in the finite temperature formalism involves the additional contributions
\begin{align}
    \frac{\partial A^\mathrm{elec}_\lambda}{\partial \lambda} \text{(2nd term)}
    = \frac{\partial A^\mathrm{elec}_\lambda}{\partial N^\mathrm{elec}} \frac{\dd N^\mathrm{elec}}{\dd \lambda}
    + \frac{\partial A^\mathrm{elec}_\lambda}{\partial T^\mathrm{elec}} \frac{\dd T^\mathrm{elec}}{\dd \lambda}
    = \mu \frac{\dd N^\mathrm{elec}}{\dd \lambda} - S^\mathrm{elec} \frac{\dd T^\mathrm{elec}}{\dd \lambda}
    \label{eq:dAdl_2}
\end{align}

\section{Implementation and Computational Details}

We have implemented the Hamiltonian interpolation approach and thermodynamic integration on top of the \textsc{PySCF} package~\cite{sun2020recent}. For reference, we also provide an implementation of the energy interpolation approach.
When the alchemical transformation does not involve an electron number change, we use zero-temperature electronic structure theory, and we have explored alchemical transformation at the Hartree-Fock and MP2 levels. We present also finite-temperature alchemical transformations at the finite-temperature Hartree-Fock level.

Below we describe additional technical computational details of the benchmark calculations and applications that will be discussed in the next section.

\subsection{Absolute protonation free energy}

The free energy change was computed at the Hartree-Fock (HF)/cc-pVDZ level and the second-order Møller–Plesset perturbation theory (MP2)/cc-pVDZ level. The nuclear charge of one of the hydronium protons was linearly scaled to zero according to $\lambda$. Simultaneously, the $\hat{h}^\mathrm{orb}$ potential on its orbitals was %progressively 
introduced to penalize electron populations on these orbitals that are absent in the pure water system. A value of $v^\mathrm{orb}=50$ Hartree for HF and 500 Hartree for MP2 was found to be adequate to ensure a small error ($\sim$0.01 kcal/mol) in the single-point energy. Concurrently, a geometric restraint was applied to the hydrogen atom, 
\begin{align}
     E^\mathrm{res}_\lambda = (1-\lambda) k^\mathrm{res} |\mathbf{R}_\mathrm{H} - \mathbf{R}_\mathrm{X}|^2
     \label{eq:h3o_res}
\end{align}
Here, $k^\mathrm{res}=0.01$ Hartree/Bohr$^2$, and $\mathbf{R}_\mathrm{X}=(2\mathbf{R}_\mathrm{O}-\mathbf{R}_\mathrm{H_1}-\mathbf{R}_\mathrm{H_2}) / 1.2 + \mathbf{R}_\mathrm{O}$, determines an approximate position of the third proton in a possible hydronium structure.

\subsection{Absolute free energy change from methanol to fluoromethane}

The free energy change was computed at the HF/3-21g level. A $v^\mathrm{orb}$ of 500 Hartree was sufficient to suppress the HF single-point energy error (with an error on the order of 0.01 kcal/mol). Instead of creating a new fluorine atom in the $\lambda=1$ state, we chose to linearly scale with $\lambda$ the nuclear charge of oxygen to that of fluorine and to simultaneously scale the charge of the hydroxyl hydrogen. Compared to the transformation that completely annihilates \ce{-OH} first, followed by creating \ce{-F}, such a transformation involves a quantum mechanically smaller perturbation and avoids changing the number of electrons. The same geometric restraint potential as in Eq.~\ref{eq:h3o_res} was used, with 
$\mathbf{R}_\text{X}=(\mathbf{R}_\text{C}-\mathbf{R}_\text{O/F})/1.4+\mathbf{R}_\text{C}$.

\subsection{Absolute free energy change from iso-butane to tert-butyl group}

This alchemical transformation changes a closed-shell molecule into an open-shell radical. We described both states with spin-symmetry-broken HF/3-21g to treat them on an equal footing. In addition to linear scaling the nuclear charge of the hydrogen atom with respect to $\lambda$, we also linearly scaled the number of $\beta$-spin electrons, keeping the $\alpha$-spin electron number unchanged and maintaining the system charge neutrality for all $\lambda$. We used finite-temperature smearing with a constant $T^\mathrm{elec}$ of 0.02 Hartree, which is small enough for the electronic entropy $S^\mathrm{elec}$ to be virtually zero ($\sim$$10^{-5}$ kcal/mol), while large enough for stable SCF convergence for the intermediate $\lambda$-states. A $v^\mathrm{orb}$ of 50 Hartree was high enough for a small error ($\sim$0.01 kcal/mol) in the HF energies. The same geometric constraint (Eq.~\ref{eq:h3o_res}) was applied with $\mathbf{R}_\text{X}=(\mathbf{R}_0-\mathbf{R}_1)\times(\mathbf{R}_0-\mathbf{R}_2)/(4~\text{Bohr})+\mathbf{R}_0$ where 0, 1, 2 represent the central carbon and two methyl carbons.

\subsection{\pka~calculations of amino acids in water}

A system of solvated lysine (Lys)/aspartic acid (Asp) in water was prepared by CHARMM-GUI\cite{jo2008charmm}. One amino acid in its protonated form with ACE and CT2 terminal capping was solvated in a TIP3P\cite{mackerell1998all,jorgensen1983comparison} water box with a $40$~\AA~side length. One chloride (as well as one potassium for the aspartic acid system) ion was added. The CHARMM36 force field\cite{best2012optimization} was used in classical equilibration and its Lennard-Jones and partial charges were used in QM/MM simulations.

The system was first energy minimized and equilibrated at the force field level for 25 ps and 50 ps in the \textit{NVT} and \textit{NPT} ensembles, respectively, both at a temperature of 298.15 K using a time step of 1 fs. The \textit{NPT} equilibration was conducted at a pressure of 100 kPa. 

The QM/MM simulations started from the last frame of the \textit{NPT} equilibration. Twenty-one $\lambda$ windows were simulated for each system with an even spacing of 0.05 and each window was run for 20 ps. The electronic structure was described at the $\omega$B97X-3c level\cite{muller2023omegab97x}. Only the side chains were treated quantum mechanically while the remaining backbone, as well as all the solvent and ions, were described classically at the CHARMM36 force field level. All QM/MM MD was performed in the \textit{NVT} ensemble at 298.15 K using a time step of 1 fs. When computing the QM--MM electrostatics, the backbone carbonyl (\ce{-CO-}) and amino (\ce{-NH-}) groups, and the $\alpha$ hydrogen, were set to have zero charges. The charges of a bonded hydrogen on the side chain and the chloride ion were linearly scaled by  $\lambda$.  The penalty potential on the ghost orbitals of the proton was set as $v^\text{orb}=50$ Hartree. A geometric restraint with the form of Eq.~\ref{eq:h3o_res} was applied with $k^\text{res}=0.1$ Hartree/Bohr$^2$ and $\mathbf{R}_\text{X}$ being the approximate position of the bonded proton. Specifically, for aspartic acid, $\mathbf{R}_\text{X}=\mathbf{R}_\text{O}+(\mathbf{R}_{\text{C}_\gamma}-\mathbf{R}_{\text{C}_\beta})/1.5$, where O is one of the carboxylic oxygens, C$_\gamma$ is the carboxylic carbon and C$_\beta$ is the $\beta$ carbon. For lysine, $\mathbf{R}_\text{X}=-0.77047501\mathbf{R}_\text{C}+3.77812947\mathbf{R}_\text{N}-1.00382723\mathbf{R}_{\text{H}_1}-1.00382723\mathbf{R}_{\text{H}_2}$, where C and N are the carbon and nitrogen in the side-chain \ce{-CH2NH3+} group and H$_1$ and H$_2$ are two of the hydrogens. When computing the \pka~difference between Lys and Asp, the free energy due to the geometric restraint cancels out because of the same $k^\text{res}$ used in both systems.

We used the QM/MM--Multipole approach\cite{li2024accurate} to treat the long-range electrostatics and $R_\text{cut}$ was updated every 5 MD steps, increasing $R_\text{cut}$ starting from 15~\AA~ with a step size of 1~\AA~until two consecutive steps both gave an octupole error below $2\times10^{-5}$ Hartree. 
The raw free energy change for the conversion of AspH + \ce{Cl-}
to Asp$^-$ + \ce{Cl}, directly computed from TI, was corrected by $k_BT\ln(2)$
to account for the entropic stabilization of the protonated Asp, which arises from the two possible proton binding carboxylic oxygens. 
Similarly, the free energy change of LysH$^+$ + \ce{Cl-} to Lys + \ce{Cl} was corrected by $-k_BT\ln(3)$, for the entropic stabilization of the deprotonated Lys due to the three possible protons that can be lost from the ammonium group. Such an entropic contribution to the free energy was not accounted for in the TI since the protonation/deprotonation step involved only one specific carboxylic oxygen and one ammonium hydrogen in the simulations.

\section{Numerical Validation}

\begin{figure}
    \centering
    \includegraphics[width=16cm]{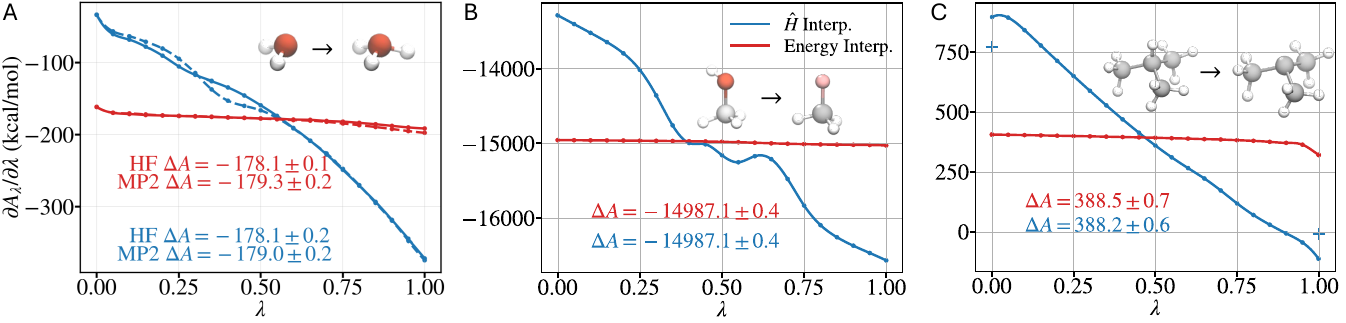} 
    \caption{The free energy gradient with respect to $\lambda$ in the alchemical transformation (A) from \ce{H2O} to \ce{H3O+}, (B) from \ce{CH3OH} to \ce{CH3F} and (C) from \ce{CH(CH3)3} to \ce{\cdot C(CH3)3}. Solid lines represent calculations at the HF level, while dashed lines represent calculations at the MP2 level. The dots show the positions of $\lambda$ windows and the sampled mean $\lambda$ gradient and the curves are the cubic spline interpolations. Blue and red lines represent Hamiltonian (single-topology) interpolation and energy (dual-topology) interpolation, respectively. The free energy change ($\Delta A=A_1-A_0$) is given in the unit of kcal/mol. In (C), the curve corresponding to Hamiltonian interpolation is plotted in the range of $\lambda=0.001$ to 0.999, and the endpoint values at $\lambda=0$ and 1 are shown by crosses.}
    \label{fig:gas-phase}
\end{figure}

We now present three examples of increasing complexity, to validate our theory and implementation. Specifically, we consider the free energy of three alchemical changes in the gas phase: (1) \ce{H2O}$\rightarrow$\ce{H3O+}, (2) \ce{CH3OH}$\rightarrow$\ce{CH3F}, and (3) \ce{CH(CH3)3}$\rightarrow$\ce{\cdot C(CH3)3}. The first two changes only involve nuclear charge changes while the third one annihilates both a nucleus and an electron. All simulations were conducted at a temperature of 310.15K and the MD integration was performed using a Langevin thermostat with a time step of 0.5 fs by i-Pi\cite{kapil2019pi}. The $\lambda$ window positions are indicated in Figure~\ref{fig:gas-phase} and each window was run for 5 ps.

In Figure~\ref{fig:gas-phase}, we show the free energy gradient ($\partial A/\partial \lambda$) profiles and the $\Delta A$ results of both the Hamiltonian and the energy interpolation approaches. We note that the gradient profiles are different between the two approaches (as expected) because they represent different alchemical paths that connect the two endpoint systems. The $\Delta A$ values, however, agree with each other within the statistical errors in all cases. This provides a basic validation of our Hamiltonian and energy interpolation implementations.

Interestingly, in the butane-to-butyl case, there are notable discontinuities in $\partial A_\lambda / \partial \lambda$ at the two endpoints using the Hamiltonian interpolation scheme. This is due to the large $\mu$ response to $N_\text{elec}$ (and thus to $\lambda$) near the two endpoints:
\begin{align}
    \frac{\partial\mu}{\partial N_e}=\frac{1}{\beta_\text{elec} \sum_i (f_i-f_i^2)}
\end{align}
when all the orbital occupations are close to either 0 or 1. The discontinuity in $\mu$ thus introduces jumps in $\partial A/\partial\lambda$ via Eq.~\ref{eq:dAdl_2}. However, jump discontinuities cannot contribute to the integral value and TI can be performed with a modified $\partial A/\partial\lambda$ whose values at $\lambda=0$ and 1 are replaced by the limit from the right and the limit from the left. In practice, we used the following gradient
\begin{align}
\frac{\partial \Bar{A}_\lambda}{\partial \lambda} = 
    \begin{cases}
        \frac{\partial A_\lambda}{\partial \lambda}  & , 0.001 \leq \lambda \leq 0.999 \\
        \frac{\partial A_\lambda}{\partial \lambda}|_{\lambda=0.001}  & , 0 \leq \lambda < 0.001 \\
        \frac{\partial A_\lambda}{\partial \lambda}|_{\lambda=0.999}  & , 0.999 < \lambda \leq 1 
    \end{cases}
\end{align}
to compute $\Delta A$ in Figure~\ref{fig:gas-phase}C. We see that the Hamiltonian and energy interpolation approaches indeed give consistent results. To further confirm that the gradient discontinuities do not contribute to the free energy change, we computed the free energy differences $A_{0.001}-A_0$ and $A_1-A_{0.999}$  using the FEP formalism, which does not involve calculations of $\partial A/\partial\lambda$:
\begin{align}
    A_{\lambda_1} - A_{\lambda_2} = -k_BT \ln{\langle \exp{\beta (E_{\lambda_2}-E_{\lambda_1})}\rangle_{\lambda_2}}
\end{align}
Comparing these results to $\partial A_\lambda / \partial \lambda \times0.001$ evaluated at $\lambda=0.001$ and 0.999, respectively, 
we find good agreement (deviations of 0.012 kcal/mol and 0.015 kcal/mol respectively).

\section{Example Application: \pka~Calculation of Amino Acids}
As a proof-of-concept application, we performed first-principle calculations of the \pka~difference between the side chains of a lysine and an aspartic acid. The two amino acids were chosen due to their large \pka~difference and different charge states in their protonated forms, imposing a greater challenge for first-principle methods (since similar chemical systems would be easily predicted to have similar p$K_\text{a}$'s by free energy calculations).

The \pka~of a weak acid is related to the free energy change of the chemical reaction (at infinitely diluted concentrations of \ce{HA} and \ce{A-}):
\begin{align}
    \ce{HA} + \ce{H2O} = \ce{A-} + \ce{H3O+}
    \label{eq:HA}
\end{align}
as
\begin{align}
    \text{p}K_\text{a} = \frac{\Delta A}{k_BT\ln{10}}-\log{\ce{[H2O]}}
\end{align}
Eq.~\ref{eq:HA} can be decomposed into two steps:
\begin{align}
    \ce{HA} + \ce{Cl-} &\rightarrow \ce{A-} + \ce{Cl} \label{eq:deprot_step}\\
    \ce{H2O} + \ce{Cl} &\rightarrow \ce{H3O+} + \ce{Cl-}
\end{align}
where a counter ion is discharged and recharged to maintain a neutral simulation box. The \pka~difference between two weak acids only involves the free energy change of the first step (Eq.~\ref{eq:deprot_step}), $\Delta A_1$:
\begin{align}
    \Delta\text{p}K_\text{a}=\frac{\Delta\Delta A_1}{k_BT\ln{10}}
\end{align}
We simulated the deprotonation of both Lys and Asp side chains and the discharging of a \ce{Cl-} by quantum mechanically annihilating the bonded proton and alchemically scaling an MM chloride charge. The resulting $\Delta\text{p}K_\text{a}=\text{p}K_\text{a}(\text{Lys})-\text{p}K_\text{a}(\text{Asp})$ is shown in Table~\ref{tab:pka}. Even at the crude level of approximation of our simulations (only the side chains are treated quantum mechanically), we find reasonable agreement ($\sim$1.7 \pka~units) between the theoretical and the experimental value (as a comparison, an error of around 1.6 \pka~units was found in an early classical constant-pH study\cite{lee2004constant}). In future studies we will explore whether quantitative agreement can be achieved by converging the simulation box, the QM region size, and the level of electronic structure theory, and by correcting for nuclear quantum effects.

\begin{table}[]
    \centering
    \begin{tabular}{cc}
    \hline
     Expt.    &   Theory \\
     \hline
     6.67$^\dagger$    &  8.4$\pm$0.3 \\
    \hline
    \end{tabular}
    \caption{\pka~difference between Lys and Asp in water.}
    \label{tab:pka}
    $^\dagger$ Taken from Ref.~\cite{asimov1960data}
\end{table}

\section{Conclusions}
We have presented a general alchemical approach for computing free energy changes between two arbitrary chemical systems at the quantum mechanical level by directly interpolating the electronic Hamiltonian. We show that the ground state of the interpolated Hamiltonian can be solved for at both an electronic mean-field and at a correlated wavefunction level, enabling alchemical free energy simulations at different levels of theory. Moreover, the derivative with respect to the alchemical parameter $\lambda$ is straightforward to compute and requires only minor changes to existing quantum chemistry software. This facilitates the computation of the free energy by thermodynamic integration.

We validated the correctness of our Hamiltonian interpolation approach against the more commonly discussed dual-topology energy interpolation formalism. We also showed a proof-of-concept application of first-principles methods to the \pka~calculation of amino acid side chains in a QM/MM setup. Our theoretical value only agrees with the experimental reference moderately well, but our general formulation allows this to be further improved in the future by converging our theoretical treatment in various aspects.

Our interpolated Hamiltonian approach requires one single quantum calculation per MD step regardless of how many alchemical transformations (i.e., how many $\lambda$'s) are simulated concurrently. This is in contrast to exponential scaling with respect to the number of $\lambda$'s in the more common energy interpolation approach. We thus view this work as the first step towards a quantum mechanical generalization of multi-$\lambda$-dynamics, which will allow for the efficient sampling of multiple alchemical changes coupled to electron phenomena.

\begin{acknowledgement}
This work was primarily supported by the US Department of Energy, Office of Science, Basic Energy Sciences, through  Award No. DE-SC0023318. GKC acknowledges additional support in the conceptualization phase from the Dreyfus Foundation, under the program Machine Learning in the Chemical Sciences and Engineering, and from the Simons Investigator program.

%Please use ``The authors thank \ldots'' rather than ``The
%authors would like to thank \ldots''.
%
%The author thanks Mats Dahlgren for version one of \textsf{achemso},
%and Donald Arseneau for the code taken from \textsf{cite} to move
%citations after punctuation. Many users have provided feedback on the
%class, which is reflected in all of the different demonstrations
%shown in this document.

\end{acknowledgement}

%\begin{suppinfo}
%\end{suppinfo}

\bibliography{quantum_ti}

\end{document}